\documentclass[a4paper,11pt]{article}
\usepackage{pos}
\usepackage[utf8]{inputenc}   
\usepackage[T1]{fontenc}      

\title{Cascades of gluons at high energies and their QI measures}

\author*[a]{Krzysztof Kutak}
\author[b]{Michał Praszałowicz}

\affiliation[a]{Instytut Fizyki Jądrowej Polskiej Akademii Nauk,\\
  Radzikowskiego 152, Krakow, Poland}

\affiliation[b]{Uniwersytet Jagielloński,\\
Łojasiewicza 11, Krakow, Poland}

\emailAdd{krzysztof.kutak@ifj.edu.pl}
\emailAdd{michal.praszalowicz@uj.edu.pl}

\abstract{In this contribution we report on recent extension of one dimensional dipole cascade models 
to account for saturation and transition to vacuum \cite{Kutak:2025syp}. 
We analyze properties of the models using Quantum Information tools. 
Furthermore, we present a description of the hadronic entropy measured by LHCb and predictions for the purity measurement.
}

\FullConference{QCD at the Extremes (QCDEX2025)\\
1–5 Sept 2025\\
Online\\}

\begin{document}
\maketitle

\section{Introduction}
Recently high-energy physics has begun to ask questions similar 
to those raised by the foundations of quantum mechanics \cite{Afik:2025ejh}.  
In particular, it has been proposed that entanglement can be studied in the process of 
Deep Inelastic Scattering (DIS) where an electron probes a proton with a virtual photon \cite{Kharzeev:2017qzs}.
This process has been studied in the high-energy limit of QCD \cite{Kharzeev:2017qzs} where it is natural to formulate it using
dipole degrees of freedom \cite{Mueller:1993rr}. The proton wave function is constructed from color dipoles. 
With this system of dipoles one can associate entropy, which is the same as the entropy of measured hadrons.
This picture has been largely confirmed by recent studies of the DIS data \cite{Hentschinski:2022evidence,Hentschinski:2022maxentDIS,Hentschinski:2024qcd_evo,Hentschinski:2023maxent}, 
and there are indications that it holds for the $pp$ case as well \cite{Tu:2019ouv,Datta:2024hpn}. 

\section{Dipole evolution equations}
\label{sec:equations}

Recent progress has been achieved by studying one-dimensional 
reduction of the equation for dipole multiplicities \cite{LevinLublinsky:2003Linear,Kharzeev:2017qzs} 
\begin{equation}
\partial _{y}p_{n}(y) =-\alpha np_{n}(y)+\alpha (n-1)p_{n-1}(y)  \, .
\label{eq:Equation0}
\end{equation}
Here, $\alpha$ corresponds to the splitting kernel, which in the one-dimensional case  is just a number. 
It directly corresponds to the Pomeron intercept. In the BFKL case 
$\alpha=4 \ln 2 (N_c\alpha_{\rm s}/\pi)$, where $\alpha_{\rm s}$ is a strong coupling constant, here, it has to be either fitted or taken as the BFKL value.

However, as the energy increases, it becomes necessary to take into account the unitarity corrections. 
In general, this is a complicated problem, leading to nonlinear evolution equations that are nonlocal in transverse dimension i.e. BK or JIMWLK equations. 
In the 1D scenario, one method is to include
terms corresponding to dipole recombination. 
The resulting equation reads \cite{Iancu:2004iy}
\begin{align}
\partial_{y} p_{n}(y) &= -\alpha n p_{n}(y) + \alpha (n-1) p_{n-1}(y) \notag \\
                      &\quad + \beta n(n+1) p_{n+1}(y) - \beta n(n-1) p_{n}(y) \, ,
\label{eq:EquationSat1}
\end{align}
where $\beta
\simeq \alpha \alpha _{\text{s}}^{2}$.
Equation (\ref{eq:EquationSat1}) models contribution corresponding to pomeron loop diagrams
and slows down  the increase of mean multiplicity.

Since, as it can be seen from the forward hadronic multiplicity data,  the multiplicity decreases 
in forward rapidities, we generalized in \cite{Kutak:2025syp}
equation~(\ref{eq:EquationSat1}) to the case where 
two dipoles can annihilate to vacuum, $2\rightarrow0$. Annihilation terms can also mimic
disappearance of particles from the detector or regions of phase space that are not available in the considered measurement.
This motivated us to add two more terms proportional to a new parameter
$\gamma$. 
The resulting equation reads as follows
\begin{eqnarray}
\partial _{y}p_{n}(y) &=&-\alpha np_{n}(y)+\alpha (n-1)p_{n-1}(y)  \notag
\\
&&+\beta n(n+1)p_{n+1}(y)-\beta n(n-1)p_{n}(y)  \notag \\
&&+\gamma (n+1)(n+2)p_{n+2}(y)-\gamma n(n-1)p_{n}(y)\, .
\label{eq:eqsat2}
\end{eqnarray}

\section{Properties of dipole cascades}
\label{sec:numcascades}

In this section, we perform numerical analysis of Eqs.~(\ref{eq:EquationSat1}) and (\ref{eq:eqsat2}).
We define parameter
$r$ that allows to express $\beta$ in terms of $\alpha$
\begin{equation}
    \beta=r \, \alpha \, .
\end{equation}

\begin{figure}[h!]
  \centering
    \centering
 \includegraphics[width=7.5cm]{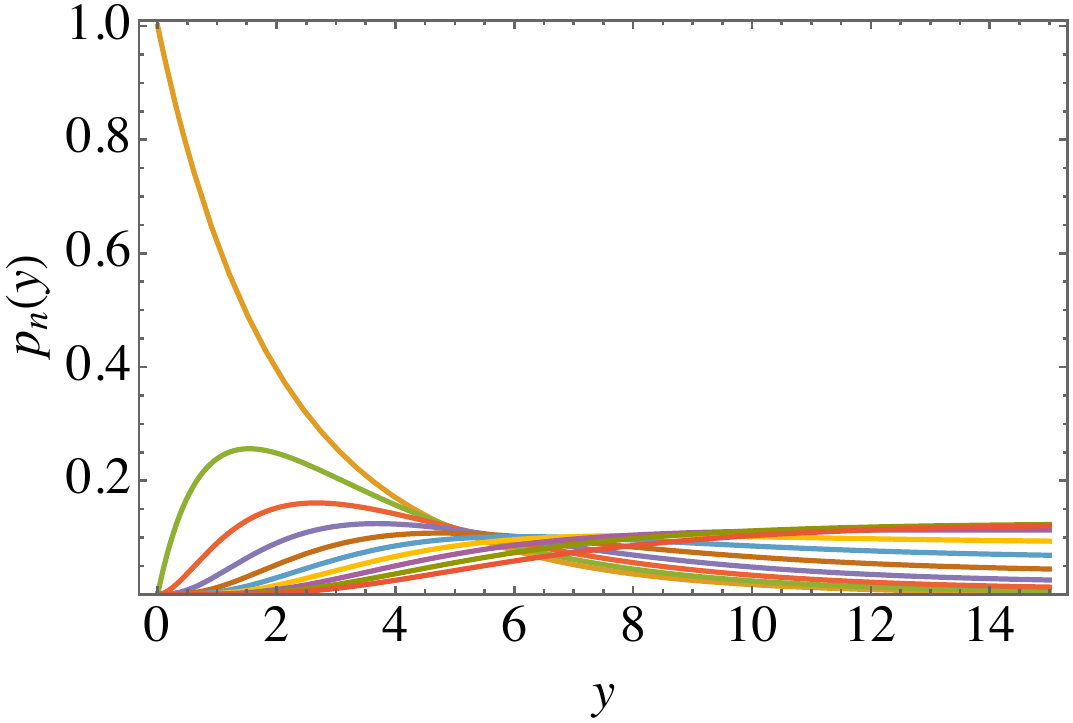}~
 \includegraphics[width=7.5cm]{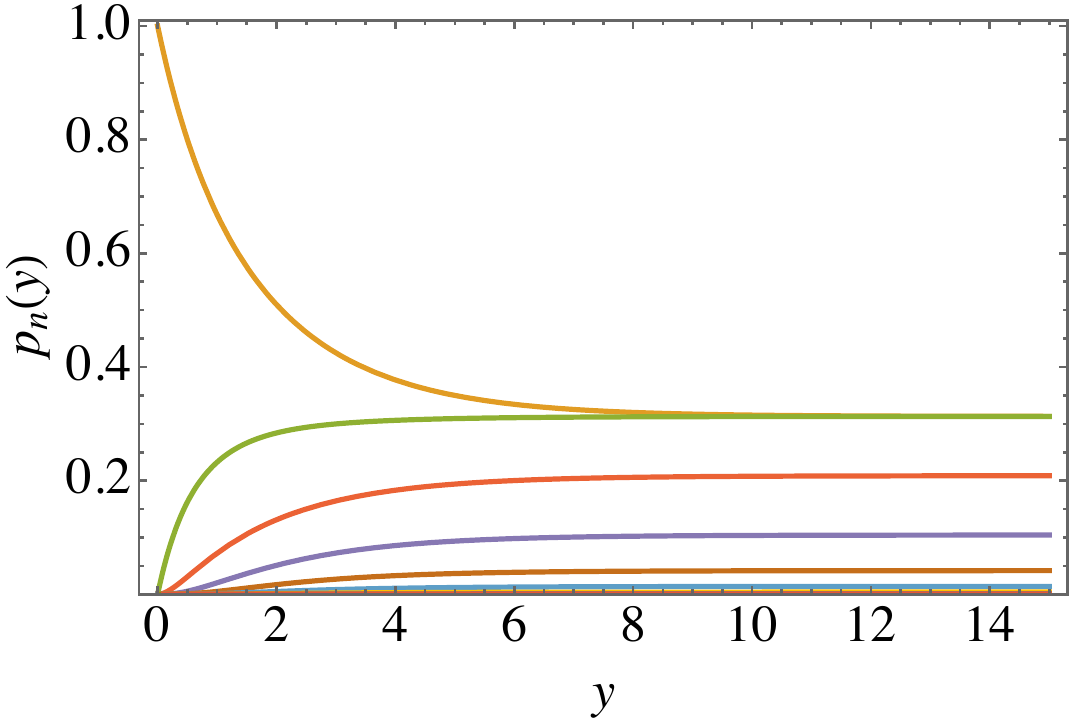}  
 \caption{Probabilities $p_n(y)$ ($n=1,\ldots ,10$) for the $\beta$ branching (\ref{eq:EquationSat1})
 for $r=0.1$ (left)  
 and $r=0.5$ (right), and $\alpha=0.5$.
 One can see that for small $r$ probabilities cross at $y_{\rm cross}\simeq 6.5$, whereas for larger $r$ they 
 reach asymptotic values without crossing.}
  \label{fig:pnyr}
\end{figure}

First, we consider Eq.~(\ref{eq:EquationSat1}), which we will call the $\beta$-cascade, in contrast to the $\alpha$-cascade
of Eq.~(\ref{eq:Equation0}) and the $\gamma$-cascade of Eq.~(\ref{eq:eqsat2}). In Fig.~\ref{fig:pnyr} we plot first 10
probabilities for $\alpha=0.5$, and $r=0.1$ and $0.5$. We see that for small
$r$ probabilities cross in the  vicinity of some rapidity, which we call $y_{\rm cross}$, and then rearrange. On the other hand, for larger $r$'s
probabilities do not cross and are almost parallel for large values of $y$. 
In what follows, we will refer to these two regimes as {\em focal} and {\em parallel}, respectively.
The physical region of $\alpha$ and $\beta$ couplings relevant to QCD evolution is in the focal regime. 
Numerical analysis allowed us to find an approximate analytical solution of  Eq.~(\ref{eq:EquationSat1}) in the
form of a Negative Binomial Distribution, which we discuss in detail in \cite{Kutak:2025syp}.

\begin{figure}[h!]
  \centering
    \centering
 \includegraphics[width=7.5cm]{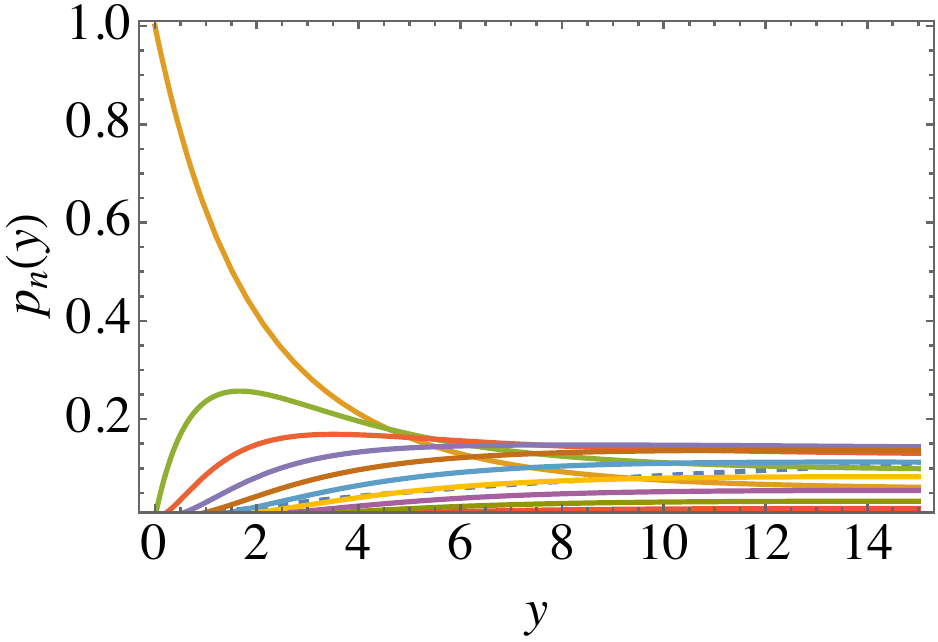}~
 \includegraphics[width=7.5cm]{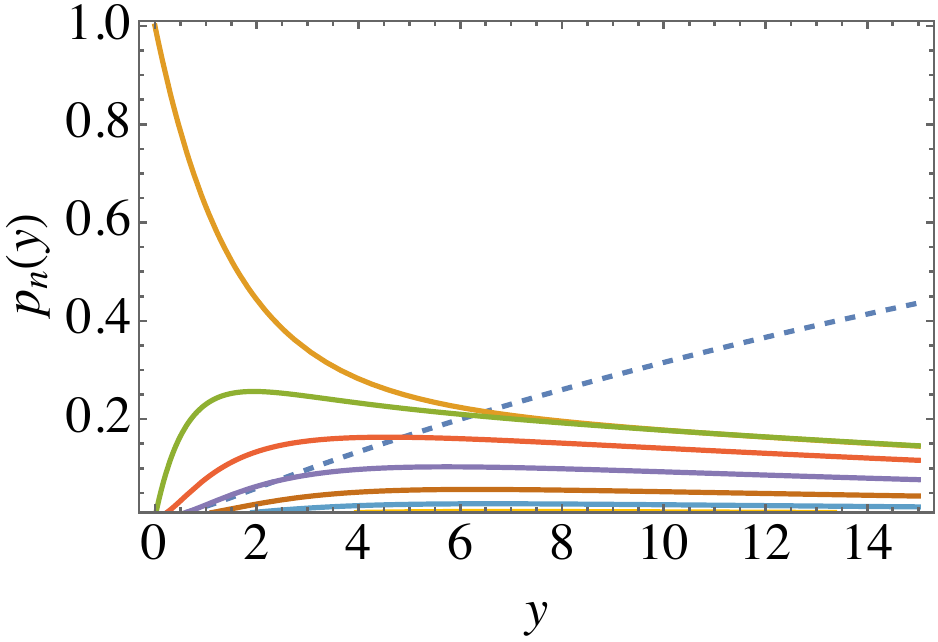}
\caption{Probabilities $p_n(y)$ ($n=0,\ldots ,10$) for the $\gamma$ branching (\ref{eq:eqsat2})
 for $\alpha=0.5,~r=0.1$ and $s=0.5$ (left),  
 and $s=1.5$ (right). Probability $p_0$ is shown as a blue dashed line.
The corresponding $\beta$-cascade is shown in the left panel of Fig.~\ref{fig:pnyr}.}
  \label{fig:pngyrs}
\end{figure}

Now we turn to Eq.~(\ref{eq:eqsat2}) describing the  cascade with recombination and with additional
$\gamma$-term ($\gamma$-cascade), which we parametrize as
\begin{equation}
   \gamma=s \beta \, . 
\end{equation}
The $\gamma$-term is responsible for the disappearance of particles
during evolution. An immediate consequence of this process is an 
increase of $p_0$ with $y$, which was identically zero in the case of cascade (\ref{eq:EquationSat1}).
Moreover, the $\gamma$ term is {\em defocusing} the probability distribution. This is shown in Fig.~\ref{fig:pngyrs}
where the $\gamma$ term with $s=0.5$ and 1.5 is added to the $\beta$-cascade shown
in the left panel of Fig.~\ref{fig:pnyr}.

Again, the numerical analysis presented above allowed us to find an approximate analytical 
solution of equation Eq.~(\ref{eq:eqsat2}) for the probability distribution excluding $p_0$
in the form of a
 Negative Binomial Distribution, which we discuss in detail in \cite{Kutak:2025syp}.


\section{Entropy other quantum measures and data description}
\label{sec:EQI}

In this section, we discuss the behavior of entropy and other quantum measures for the
cascades studied in the previous section. 
The von Neuman entropy of reduced density matrix that follows from integrating  out all degrees of freedom except rapidity \cite{Kharzeev:2017qzs,Liu:2022hto} 
 reads
\begin{equation}
S(y) = -\mathrm{Tr}\,\rho \ln \rho 
     = - \sum_{n} p_n(y) \, \ln\!\big( p_n(y) \big) \, ,
\label{eq:Sdef}
\end{equation}
where $p_n(y)$ are interpreted as probabilities for a given number of dipoles at rapidity $y$. 
It measures how much information has been lost due to integrating out quarks and averaging over transverse 
and  color degrees of freedom, as evolution in rapidity progresses. 

In addition to entropy, solutions of the cascade equations can be used to calculate various  quantum information (QI) measures. 
In particular, the mean number of dipoles $\bar{n}$, called in the QI context a complexity, measures how the underlying 
quantum state spreads in the Hilbert space.
The variance quantifies the fluctuations around the average. Here we use the normalized variance 
defined as follows
\begin{equation}
    \delta^2=\frac{\langle n^2\rangle-\langle n\rangle^2}{\langle n\rangle^2} \, .
\end{equation}
Finally, the quantity that measures the purity, {\em i.e.}
the extent to which a system deviates from a pure state as evolution progresses, is defined as
\begin{equation}
    \gamma=\sum_n p_n^2 \, .
    \label{eq:purdef}
\end{equation}

\begin{figure}[t!]
  \centering
  \begin{minipage}[b]{0.45\textwidth}
    \centering
    \includegraphics[width=\linewidth]{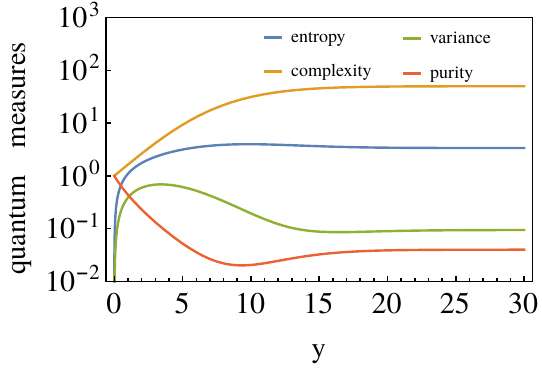}
  \end{minipage}
 \begin{minipage}[b]{0.45\textwidth}
    \centering
    \includegraphics[width=\linewidth]{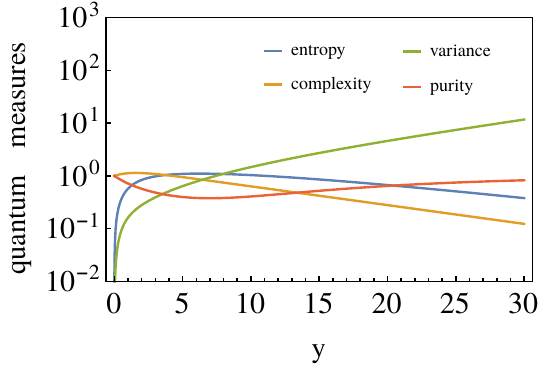}
  \end{minipage}\hfill
  \caption{ In the left panel: quantum measures as obtained from solutions of $\beta$-cascade Eq.~(\ref{eq:EquationSat1}). 
  In the right panel: quantum measures as obtained from solutions of $\gamma$-cascade Eq.~(\ref{eq:eqsat2}).}
  \label{fig:results22}
\end{figure}

In the following, we 
compute these Quantum Measures for the $\beta$- and $\gamma$-cascades, which is illustrated in Fig.~\ref{fig:results22}. 

\begin{figure}[t!]
  \centering
  \begin{minipage}[b]{0.45\textwidth}
    \centering
    \includegraphics[width=\linewidth]{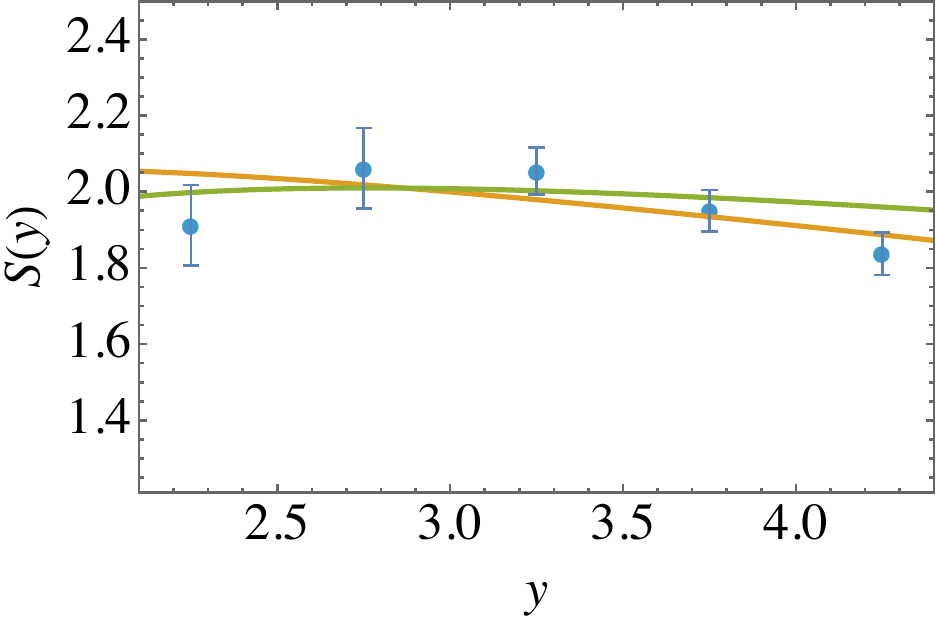}
  \end{minipage}~~
 \begin{minipage}[b]{0.45\textwidth}
    \centering
    \includegraphics[width=\linewidth]{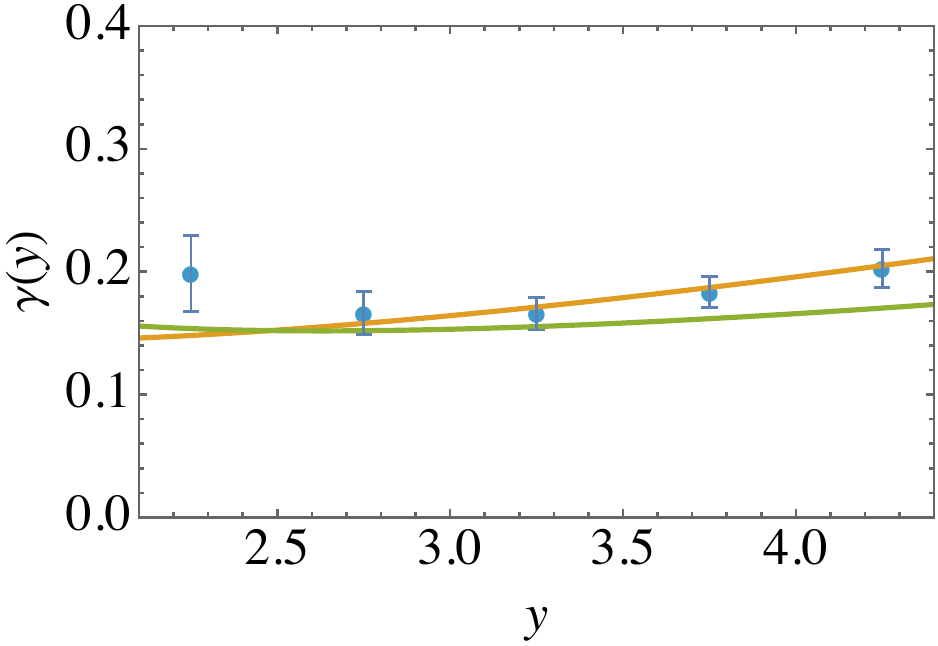}
  \end{minipage}\hfill
  \caption{Entropy (left panel) and purity (right panel)  obtained from the $\gamma$-cascade (solid lines)
  vs. data. Parameters of the $\gamma$-cascade
  were obtained
  by minimizing the $\chi^2$ for entropy (see text).
  Purity is therefore a prediction.}
  \label{fig:EPplot}
\end{figure}

With the developed framework we address  description of hadronic entropy and provide prediction for purity. 
The entropy and purity data can be derived from probability distributions measured by the LHCb  collaboration \cite{LHCb:2014wmv}.
The results are shown in Fig.~(\ref{fig:EPplot}).

\newpage
\bibliographystyle{jhep} 
\bibliography{references.bib}

\end{document}